\documentclass{emulateapj}

\def\brfrac#1#2{\left(\dfrac{#1}{#2}\right)}
\def\dfrac#1#2{{\displaystyle\frac{\mathstrut #1}{#2}}}

\slugcomment{Received 2010 October 6; accepted 2010 October 25}

\shorttitle{Orientation Effects on AGN Dusty Torus}
\shortauthors{Kawaguchi and Mori}

\begin{document}

\title{Orientation Effects on the Inner Region of 
Dusty Torus of Active Galactic Nuclei}

\author{Toshihiro Kawaguchi and Masao Mori}
\affil{Center for Computational Sciences, University of Tsukuba, 
Tsukuba, 
Ibaraki 305-8577, Japan}

\email{kawaguti@ccs.tsukuba.ac.jp}

\begin{abstract}
A sublimation
process governs the innermost region of the dusty 
torus 
of active galactic nuclei.
However, the 
observed inner radius of the torus is 
systematically smaller than the expected radius 
by a factor of $\sim 1/3$.
We show that the anisotropy of the emission 
from accretion disks resolves this 
conflict naturally and quantitatively.
An accretion disk emits lesser radiation in the 
direction closer to its equatorial plane 
(i.e., to the torus).
We find that the anisotropy makes the torus inner region 
closer to the central black hole and concave.
Moreover, the innermost edge of the torus may connect 
with the outermost edge of the disk continuously.
Considering the anisotropic emission of each clump in 
the torus, we 
calculate the near-infrared 
flux variation 
in response to a UV flash.
For an observer at 
the polar angle 
$\theta_{\rm obs} = 25$\degr, 
the centroid 
of the time delay is found to be 37\% of the delay 
expected in the case of isotropic illumination, 
which explains the observed systematic deviation.
\end{abstract}


\section{Introduction}
Active galactic nuclei (AGNs) are powered by gas accretion onto
supermassive black holes (BHs) at the center of each galaxy.
The central BH and the accretion disk are hidden 
 from our line-of-sight, in some geometry,  
  by 
an optically- and geometrically-thick dusty torus 
 (Antonucci \& Miller 1985; 
 Miller \& Goodrich 1990). 
Since the torus potentially plays a role of a gas reservoir 
for the accretion disk, 
its nature 
has long been studied 
(Pier \& Krolik 1992;  
Efstathiou \& Rowan-Robinson 1995; 
Beckert \& Duschl 2004; 
Schartmann et al. 2008). 

A large geometrical thickness of the torus 
is inferred from the detection of high optical polarization,
the sharp-edged ionization cone,  
the near and mid-infrared (NIR \& MIR) to UV luminosity ratio
 and
the number ratio between type-1 and type-2 AGNs 
in the unified AGN scheme 
(Antonucci 1993; 
 Wilson \& Tsvetanov 1994; 
 Lawrence 1991).
To sustain the geometrical thickness of the torus, 
the vertical velocity dispersion must be 
as large as $\sim 100 \mathrm{km\, s}^{-1}$.
However, it is impossible to explain such a large velocity 
dispersion by thermal velocity, since dust grains cannot 
survive over a sublimation temperature $T_{\rm sub} \sim 1500$K 
(Barvainis 1987; Laor \& Draine 1993).
Therefore, Krolik \& Begelman (1988) concluded that 
numerous dusty clumps, 
rather than a smooth mixture of gas and dust, 
with a temperature 
$\lesssim 1500$~K 
constitute the torus with a large clump-to-clump velocity 
dispersion (i.e., dusty clumpy torus).
Various models for the clumpy torus have been 
investigated 
(Nenkova et al. 2002, 2008; 
Wada \& Norman 2002; 
Dullemond \& van Bemmel 2005; H\"{o}nig et al. 2006). 

The dusty and clumpy torus absorbs optical/UV radiation 
from the central accretion disk, and re-emits as 
IR radiation (Telesco et al. 1984; Radovich et al. 1999).
The energy balance (of each clump) between heating by incident radiation flux 
and cooling via blackbody radiation indicates that 
clumps located closer to the central BH have higher temperature.
Clumps at the innermost region of the torus have 
the highest temperature, with a temperature of $T_{\rm sub}$ at 
the irradiated surface, and emit 
NIR radiation as 
"3$\mu$m bump" 
(Kobayashi et al. 1993).
Barvainis (1987) derived the innermost radius of the torus
(dust sublimation radius, denoted as $R_{\rm sub,0}$ in this
study):  
\begin{equation}
R_{\rm sub,0} = 0.13 \brfrac{L_{\rm UV}}{10^{44} \, erg/s}^{0.5}
 \brfrac{T_{\rm sub}}{1500 K}^{-2.8}
 \brfrac{a}{0.05 \mu m}^{-0.5} {\rm pc},
 \label{eq:bar87}
\end{equation}
where $L_{\rm UV}$ 
and $a$ are 
UV luminosity 
and the size of dust grains, respectively.
If this estimation is correct, 
the NIR radiation flux varies with 
a time lag of some months behind the 
optical/UV flux variations.

Photometric monitoring observations of type-1 AGNs 
revealed that NIR emission 
from AGNs indeed lags behind optical variation by an order 
of a month
(Clavel et al. 1989; Glass 2004; 
Minezaki et al. 2004). 
Moreover, the luminosity dependency of the time lag also 
coincides with the theoretical prediction as $\propto L_{\rm UV}^{0.5}$
(Suganuma et al. 2006; 
Gaskell, Klimek \& Nazarova 2007 using the data presented 
by Glass 2004).
These agreements between the observational results and the theoretical 
considerations support the idea that the innermost radius of the 
torus is controlled by the dust sublimation caused by irradiation
from the accretion disk.

Despite the successful agreement of luminosity dependency
of the NIR-to-optical time lag, however, 
a conflict appeared concerning the normalization of the time lag. 
Namely, the time lag measured and collected by Suganuma et al. (2006) 
is systematically smaller than the lag predicted from 
eq. \ref{eq:bar87} by a factor of $\sim 1/3$ 
(Kishimoto et al. 2007; Nenkova et al. 2008).

In this {\it Letter}, we notice that 
the estimation of dust sublimation radius (eq. \ref{eq:bar87}) 
presumes isotropic emission from the accretion disk.
Yet, emission from an optically-thick disk is, in principle, anisotropic, 
which is a fact missing in previous theoretical considerations.
There is a systematic difference between the inclination angle 
at which we observe the disk in type-1 AGNs and the angle 
at which an aligned torus observes the disk.
We show that 
the effects of the anisotropic 
illumination flux from the accretion disk 
resolve
the systematic discrepancy between the observations and 
the theory regarding the NIR-to-optical time lag.
In the next section, the anisotropy of the disk emission 
is introduced. 
Then, 
the inner structure of the torus is examined 
within the framework of clumpy torus models 
in \S 3.
In \S 4, we derive the transfer function for NIR emission 
of the torus in response to optical/UV flux variation in the disk. 
Finally, we make a summary and discussion of this study in \S 5.

\section{Anisotropic Emission of Accretion Disk}

Now, let us consider the emission from  
an optically-thick, plain slab (i.e., disk).
Radiation flux ($F$) from 
 a unit surface area of the disk toward a unit 
 solid angle at the polar angle of $\theta$ 
 decreases with an increasing $\theta $ as follows:
\begin{equation}
F \propto \cos \theta \, (1 + 2 \cos \theta )
 \label{eq:flux}
\end{equation}
Here, the first term represents the change in the projected 
surface area, while the latter represents the limb darkening 
effect for plasma, whose opacity is dominated by 
electron scattering over absorption 
(Netzer 1987 and references therein).
In other words, an accretion disk emits lesser 
radiation in the direction closer to its equatorial plane 
(i.e., larger $\theta $; 
Laor \& Netzer 1989; Sun \& Malkan 1989). 
Consequently, the assumption of isotropic emission from accretion disks 
(e.g., eq. \ref{eq:bar87}) obviously overestimates the radiation flux toward the 
torus. 
Therefore, it overestimates the inner radius of the torus. 

We note that this effect works even if the disk 
is infinitesimally thin.
The fact that the disk has a nonzero thickness brings about another 
anisotropy of illumination flux, 
such that the torus is not 
illuminated below the disk height (near the equatorial plane)
at $\theta $ larger than a critical angle $\theta_{\rm max} $.
Throughout this {\it Letter}, we adopt a thin disk (with 
an aspect ratio of $\sim 0.01$), like 
the standard accretion disk model (Shakura \& Sunyaev 1973).
The effects of varying 
disk thickness  
as a function of accretion rates 
(Abramowicz et al. 1988; 
Fukue 2000; 
Kawaguchi 2003) 
will be investigated 
in a forthcoming paper.

The effect of anisotropic emission
 has already been discussed and referred to 
  as the orientation effect 
 in the context of  
emission lines, such as the Baldwin effect  
(Netzer 1985; Francis 1993; Bottorff et al. 1997).
However, consequences of this effect upon 
the torus have not  been examined so far.
Hereafter, we will explore how this effect influences the 
torus structure.

\section{Inner Structure of Dusty Torus}

\begin{figure}
\plotone{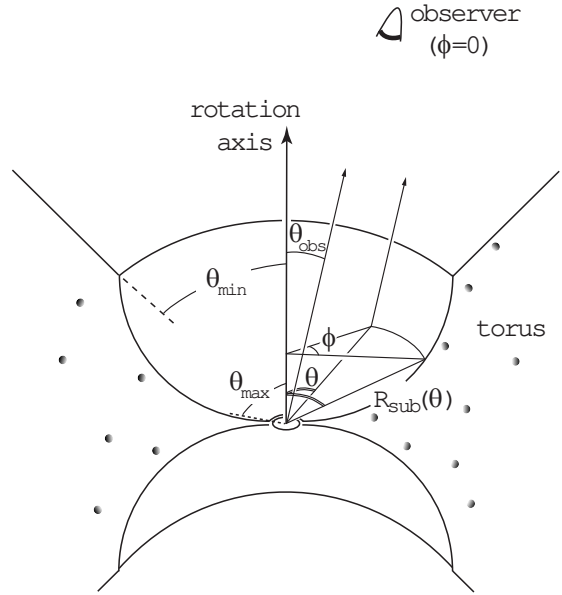}
\caption{Schematic view of the inner structure of the torus. 
On the left side of the rotational axis, 
the definitions of 
$\theta_{\rm min}$ (the torus thickness) and 
$\theta_{\rm max}$ (the disk thickness) are indicated.
\label{fig:schematic}}
\end{figure}

We determine the inner edge of the torus so that 
the temperature of a clump (at the irradiated surface) equals 
to the sublimation temperature 
at the edge.
As mentioned above, radiation flux from the accretion disk varies
with the polar angle $\theta $.
Thus, the sublimation
radius of the torus is also a function of 
$\theta $, which is indicated by $R_{\rm sub}(\theta )$ 
in this study (Figure~\ref{fig:schematic}).
Namely, 
$R_{\rm sub}(\theta )$ 
is the distance between the torus edge and the central BH 
for various $\theta $.
To avoid confusion, we express the sublimation 
radius estimated under the isotropic emission assumption (i.e., 
eq. \ref{eq:bar87}), in contrast, as $R_{\rm sub,0}$.

Adopting the anisotropic illumination given in eq. \ref{eq:flux}, 
we obtain
\begin{equation}
R_{\rm sub}(\theta ) = R_{\rm sub,0} 
\left[\dfrac{\cos \theta \, (1 + 2 \cos \theta )}
{\cos \theta_{\rm obs} \, (1 + 2 \cos \theta_{\rm obs} )}\right]^{0.5}.
\label{eq:rsub}
\end{equation}
Here, $\theta_{\rm obs}$ is the polar angle toward the 
observer seen from the central accretion disk, 
and $\theta_{\rm obs} = 25$\degr\ is assumed throughout this study.
Although various grain sizes result in the sublimation 
process occurring over a transition zone rather than a single 
distance (Nenkova et al. 2008), we employ a sharp boundary 
for simplicity.

\begin{figure}
\epsscale{.80}
\plotone{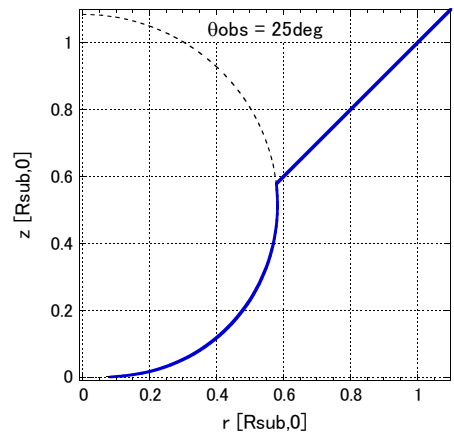}
\caption{
Calculated geometry of the innermost region of the 
dusty torus, where $r$ and $z$ are the cylindrical coordinates.
The central BH and the accretion disk in the $z=0$ plane are located at the 
origin (not drawn) and are observed with 
the polar angle $\theta_{\rm obs} = 25$\degr.
We assume alignment between the disk and the torus.
In the case of an isotropic illumination, the torus edge 
has been supposed to stand at $r=R_{\rm sub,0}$.
Thick solid line indicates the edge of the torus, 
with the opening angle of the torus ($\theta_{\rm min}$) 
assumed to be 45\degr.
On the right-hand side of the line, 
clumps contain dust.
Dashed line represents $R_{\rm sub}(\theta )$ (eq. \ref{eq:rsub}).
The torus inner edge is located closer to 
the central BH than in the previous estimations (eq. \ref{eq:bar87}),
and 
concave/hollow.  
The innermost region of the torus may connect with 
the outermost radius of the disk.
\label{fig:geometry}}
\end{figure}

Figure~\ref{fig:geometry} shows the calculated structure 
of the innermost region of the torus.
The central BH and the accretion disk on the $z=0$ plane are located at the 
origin of the coordinate axes (not drawn).
If the torus is indeed a reservoir of gas for the disk, 
angular momenta of the infalling gas will 
align these axes.
Thus, we assume alignment between the disk and the torus.
In the case of an isotropic emission from the disk, 
the torus edge has been supposed to stand at $r=R_{\rm sub,0}$.
Thick solid line indicates the edge of the torus, 
with the opening angle of the torus ($\theta_{\rm min}$) 
assumed to be 45\degr, and the maximum $\theta$ of the torus 
($\theta_{\rm max}$) set to 89\degr\ (i.e., thin disk approximation
 mentioned in the previous section).
On the right-hand side of the line, 
the temperatures of clumps are 
below the dust sublimation temperature.
Dashed line represents $R_{\rm sub}(\theta )$ (eq. \ref{eq:rsub})
and is drawn as a guide for various $\theta_{\rm min}$.
It turns out that (i) the torus inner edge is located closer to 
the central BH than suggested by previous estimations (eq. \ref{eq:bar87})
and that (ii) the structure of the edge is concave/hollow.
These are the results of weaker illumination flux toward 
larger $\theta $.
Namely, dust can survive closer to the central BH at 
larger $\theta $.

Moreover, (iii) $R_{\rm sub}(\theta )$ decreases down to 
$0.1 \times R_{\rm sub,0}$ at $\theta =88.5$\degr.
This radius coincides with the outermost radius of AGN accretion 
disks, which is determined by the onset of radial  
self-gravity of the disk (see Kawaguchi, Pierens \& Hur\'e 2004).
Little is known about the region outside the outermost radius 
of the disk (e.g., Collin 2001), such as what happens 
in the 1-dex gap between the disk outermost radius and 
the torus innermost radius ($R_{\rm sub,0}$).
Our result indicates that there is no gap between the torus 
and the disk (Emmering, Blandford \& Shlosman 1992; Elitzur \& Shlosman 2006).

\section{Transfer Function}

Although interferometric NIR observations have been made for 
nearby Seyfert galaxies, only the visibility data, rather than 
images, are achieved 
because of  
the poor $u$-$v$ coverage 
(e.g., Kishimoto et al. 2009).
Even with the next generation telescope such as 
Thirty Meter Telescope having $0.01"$ spatial resolution at NIR, 
the innermost radius of the torus discussed above cannot be spatially resolved.
Thus, observations of time variability will continue to be 
powerful tools to explore the innermost structure of the torus 
in the coming decade.
In this section, we calculate the 
time variation of NIR emission in response 
to a $\delta$-function like variation of the 
irradiation optical/UV flux (the transfer function). 

Transfer functions $\Psi (t)$ for various geometries of 
the re-emitting region have long been studied, mainly 
in the context of the broad emission line region of AGNs 
(e.g., Netzer 1990).
For instance, a ring produces a double-horned $\Psi (t)$, 
while a thin (or thick) shell gives a rectangle (or trapezoid).
Time variation of the reprocessed radiation (NIR in this study) 
is a convolution of the illumination flux variation with 
$\Psi (t)$.
The measured time lag corresponds to the centroid of $\Psi (t)$.
Below, we calculate $\Psi (t)$ and its centroid in order to 
compare them with the observational results.

It turned out that it took $\sim 1$~year for the inner region 
of the dusty torus to adjust to the varying illumination 
flux (Koshida et al. 2009; Pott et al. 2010).
We hereafter regard the inner structure of the torus 
as time-independent 
in the timescale of NIR-to-optical time lag ($\sim$months).

\subsection{Calculation Method}

To calculate $\Psi (t)$ for the clumpy torus, we discuss the following 
three items;
(a) the optical path, 
(b) NIR emissivity of the torus inner region 
and (c) anisotropic emission of each clump.
In this {\it Letter}, we ignore the response from the torus edge 
at $\theta > \frac{\pi }{2}$, assuming torus 
self-occultation (absorption of NIR emission from a clump 
by other clumps in the line of sight). 
Although we do not go further, 
this assumption can be tested in principle via 
the profile of broad emission lines and its time variation 
(e.g., Peterson 2001).

Firstly, (a) the optical path difference is written as
\begin{equation}
R_{\rm sub}(\theta ) 
\left[
1 - \{\cos \theta_{\rm obs} \, \cos \theta
 + \sin \theta_{\rm obs} \, \sin \theta \, \cos \phi \}
\right], 
\label{eq:opd}
\end{equation}
where $\phi$ is the azimuthal angle and defined so that 
$\phi = 0$ for the observer (Figure~\ref{fig:schematic}).
The concave shape of the inner region of the torus reduces the 
optical path difference.
Clumps at a distance slightly larger than $R_{\rm sub}(\theta)$
with a slightly lower temperature than the sublimation temperature will 
somehow emit NIR radiation.
Moreover, clumps at a distance slightly smaller than $R_{\rm sub}(\theta)$, 
in the middle of the dust destruction process but with some 
of their dust grains still surviving, will also emit NIR radiation.
These effects will smear out the resultant NIR response $\Psi (t)$.
Since these effects 
are unlikely to change the centroid of $\Psi (t)$ drastically, 
 we here ignore them and consider only optical paths that 
hit the inner edge of the torus.

Next, (b) we discuss the emissivity of the torus inner region 
as a function of $\theta $.
Since $R_{\rm sub}(\theta)$ varies with $\theta$, 
the distance dependency of 
the NIR emissivity must be determined. 
Following the calculations of three-dimensional radiative transfer 
(e.g., H\"{o}nig et al. 2006; Schartmann et al. 2008), 
we assume that 
(b-1) the clump size increases and 
(b-2) the clump number density  decreases 
when the clump-to-BH distance increases, as follows.

Closer to the BH, the tidal force is stronger, thus  
only smaller clumps are likely to survive there.
Therefore, we assume that the clump size is in proportion to the 
distance between the clump and the central BH.
Then, the apparent angular size (observed from the BH) of 
the clumps at the 
torus inner edge, which are emitting NIR radiation, 
is independent of $\theta$.
(Farther clumps receiving direct emission from the disk as well 
are not of interested in this study, because they emit MIR rather 
than NIR.)

Little is known about the number density of clumps.
We assume that the number of clumps at the torus edge 
per unit solid angle ($d \Omega $) is independent of $\theta$.
Namely, the same number of NIR-emitting clumps is contained 
per $d \Omega $ for various $\theta$.
To summarize the two factors, 
the emissivity of NIR flux per $d \Omega $ is 
assumed to be proportional to $R_{\rm sub}(\theta)^2$.

Lastly, (c) the anisotropy of NIR emission from each clump is considered.
To calculate NIR flux, we must care about how extent 
the illuminated surface of a clump is seen by the observer (Nenkova et al. 2002).
H\"{o}nig et al. (2006) mentioned that this effect resembles the 
phases of the moon.
Let us suppose that an observer looks at a clump with 
an angle $\xi $, where $\xi =0$ means a face-on view of the illuminated 
surface (the angle denoted as $\phi $ in the paper by H\"{o}nig et al.).
In the case of the moon phase, the anisotropy 
coefficient (the fraction of the illuminated surface seen by the 
observer) is $(1 + \cos \xi )/2$.
However, incident photons can somehow penetrate gaseous clumps, 
heating up 
the limb of each clump.
Then, the anisotropy is weakened compared with the 
above estimation for the solid (moon).
We adopt the following anisotropic coefficient for 
the waning effect, 
\begin{equation}
\min \left[
1, \left(\frac{1 + \cos \xi}{2} + 0.1 \right)
\right].
\label{eq:waning}
\end{equation}
This coefficient is chosen so as to reproduce the Monte Carlo calculations 
by H\"{o}nig et al. (2006) for a single clump observed from 
three different $\xi $.
We also try other descriptions for the anisotropy under a condition 
that each description reproduces the Monte Carlo results, 
obtaining little difference in the resultant $\Psi (t)$.

\subsection{Result}

\begin{figure}
\epsscale{.80}
\plotone{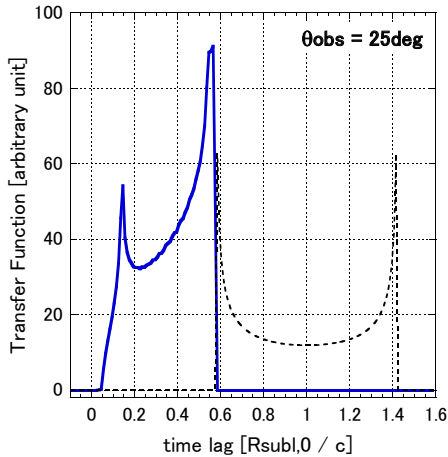}
\caption{Resultant transfer function for the torus we consider 
(solid line; 
NIR response to a $\delta$-function like variation of illumination 
flux).
Dashed line is the transfer function for a ring 
with radius $R_{\rm sub,0}$ and $\theta_{\rm obs} = 25$\degr.
In the case of the anisotropic disk emission, 
the torus responds approximately three times faster than the ring does 
(the centroid of the transfer function is 0.37$R_{\rm sub,0}/c$).
\label{fig:tf}}
\end{figure}

Figure~\ref{fig:tf} presents the resultant transfer function 
of the torus for $\theta_{\rm obs}$ = 25\degr\ (solid line), 
calculated by integrating $\theta$ (from $\theta_{\rm min}$ 
to $\theta_{\rm max}$) and $\phi $ (0 to $\pi$). 
The portion at $\phi \sim 0$ produces the left horn, 
which is  a faster 
(due to a shorter optical path difference, eq. \ref{eq:opd})
and weaker 
(due to the waning effect, \S4.1) response than 
the opposite area at $\phi \sim \pi$ (right horn).
Dashed line is the transfer function for a ring 
with radius $R_{\rm sub,0}$ 
observed from $\theta_{\rm obs}$ = 25\degr   
and is drawn 
for an example of transfer functions in the case of  
isotropic illumination flux.
The ring responds to the illumination flux at around $R_{\rm sub,0}/c$
(where $c$ is the speed of the light) after the $\delta $-function
like variation in the illumination flux at zero time lag.
(It is just a coincidence that the ring starts to respond 
exactly when the torus ends with the current parameter sets.)

On the other hand, the torus we consider shows a faster response 
and narrower width of the transfer function.
These results are due to the facts that the torus inner region 
we consider  
is (i) closer to the central BH than $R_{\rm sub,0}$ 
and (ii) concave.
The centroid of the time delay of the transfer function 
is found to be $0.37 R_{\rm sub,0}/c$, 
which solves the puzzle of the systematic deviation of a factor of $\sim 1/3$
between the observational results and the theoretical considerations 
(under the assumption of isotropic emission).

\section{Summary and Discussion}

The dusty clumpy torus surrounds the accretion disk and BH in AGNs.
Various observations have revealed that the dust sublimation
process governs the innermost region of the torus.
However, there was a systematic deviation between the observational 
results and the theory
regarding the inner radius of the torus.

In this study, we have shown that the anisotropy of the emission 
from accretion disks resolves this 
conflict naturally and quantitatively.
Namely, the angle at which we observe the disk in type-1 AGNs 
is systematically 
closer to a pole-on view than the angle at which an aligned torus 
observes the disk.
An accretion disk emits lesser radiation in the 
direction closer to its equatorial plane (larger $\theta $).
We have found that the anisotropy makes the torus inner region 
closer to the central BH and concave.
Furthermore, the innermost edge of the torus may connect 
with the outermost edge of the accretion disk continuously.

Considering the anisotropic emission of each clump, we have 
calculated the NIR flux variation of the torus in response 
to a UV flash.
A rapid response to the illumination flux is realized, 
due to the small inner radius and concave shape of the 
torus. 
For an observer at $\theta_{\rm obs} = 25$\degr, the centroid 
of the time delay is found to be 37\% of the delay 
expected in the case of isotropic illumination.

Other than the orientation effect examined in this study,
large grain size (see eq. \ref{eq:bar87}) and/or extinction 
between the torus and the accretion disk (i.e., in the broad 
emission line region) are possible concepts to reduce 
the inner radius of the torus (Laor \& Draine 1993; 
Maiolino et al. 2001; 
Gaskell et al. 2007).

Anisotropy is an inevitable feature of the emission from 
optically-thick disks.
Even if a BH object (seen in nearly face-on geometry) 
looks very luminous around the Eddington luminosity,
 the 
 flux 
 in the directions close to the equatorial plane
 is not necessarily huge to prevent gas infall. 
This effect 
(along with the disk self-occultation)
enables  
gas supply to 
luminous BH objects. 
Actually, the observed data do not support the concept of 
Eddington-limited accretion (Collin \& Kawaguchi 2004).

Future studies will examine misalignments between the torus and the disk, 
and different 
$\theta_{\rm obs}$ (for type 1.5, 1.8 etc objects), 
$\theta_{\rm min}$ (for objects with a thick/thin torus) 
and $\theta_{\rm max}$ (for thick disks 
with super-Eddington accretion rates).
For instance, 
some Narrow-Line Seyfert 1 galaxies 
and Narrow-Line QSOs prohibit weak NIR emission  
(Rodr{\'{\i}}guez-Ardila \& Mazzalay 2006;
Kawaguchi et al. 2004; 
Jiang et al. 2010; however, see Hao et al. 2010).
Small $\theta_{\rm max}$ due to the disk self-occultation 
with super-Eddington accretion rates (Fukue 2000)
can be a reason for the weakness. 

\acknowledgments

We thank 
Takeo Minezaki and
Shintaro Koshida 
for useful discussions, 
and the anonymous referee for comments.
This work was partly supported by the Grants-in-Aid of the Ministry
of Education, Science, Culture, and Sport (19740105, 21244013).

\end{document}